# Scaling of resistivities and guided vortex motion in MgB$_2$ thin films

P Vašek

Institute of Physics ASCR, Na Slovance 2, 182 21 Prague 6,
Czech Republic

E-mail: vasek@fzu.cz

**Abstract.** Longitudinal and transverse voltages have been measured on thin films of MgB$_2$ with different superconducting transition widths. The study has been performed in zero and non-zero external magnetic fields. The non-zero transverse voltage has been observed in close vicinity of the critical temperature in zero external magnetic field, while further away from T$_c$ this voltage becomes zero. In magnetic field it becomes a transverse voltage which is an even function with respect to the direction of the field. The usual Hall voltage starts to appear with increasing magnetic field and in large fields the even voltage disappears and only the Hall voltage is measurable (i.e. the transverse even voltage is suppressed with increasing magnetic field and increasing transport current). New scaling between transverse and longitudinal resistivities has been observed in the form $\rho_{xy} \sim d\rho_{xx}/dT$. This correlation is valid not only in the zero magnetic field but also in nonzero magnetic field where transverse even voltage is detected. Several models trying to explain observed results are discussed. The most promising one seems to be guided motion of the vortices, though further theoretical work will be required to confirm this.

## 1. Introduction

Since the discovery of superconductivity in magnesium diboride, scientists have devoted large attention to this superconductor because of its properties and its potential in technological applications as well. The advantages of using MgB$_2$ for technological applications are its simple crystal structure, the relatively high critical temperature and the large coherence length, which makes the materials less susceptible to structural defects like grain boundaries.

MgB$_2$ turned out to be a standard Bardeen–Cooper–Schrieffer-like (BCS-like) superconductor. The critical temperature T$_c$ is almost 40 K for this compound. Relatively high critical temperature in comparison with conventional low T$_c$ superconductors and lower anisotropy and higher stability of superconducting properties than in cuprates make MgB$_2$ superconductors a good compromise between low and high temperature superconductors. Many effects have been observed which are similar to those detected in high temperature superconductors. One example is the change of sign of the Hall effect in the mixed state just below T$_c$. Although numerous theoretical attempts have been made, adequate explanation of this observation was not reached yet.

MgB$_2$ belongs to type II superconductors with all the implications resulting from vortex structure in the mixed state. The vortices should move in the direction of the Lorentz force due to transport current. If the superconducting system has some inhomogeneities which can prevent the movement of vortices in the Lorentz force direction, the vortices can move in direction which does not coincide with that of Lorentz force. Such situation can be identified by observing an even (symmetric in field) transverse voltage which, as opposed to the odd (antisymmetric) Hall voltage, does not change sign upon magnetic field reversal [1].

However non-zero transverse voltage can be also observed in zero external magnetic field near critical temperature [2]. In this article we would like to describe experimental results observed namely the transformation of transverse voltage in zero field to the transverse even voltage in magnetic field. It should be mentioned that we will not take care about Hall voltage, i.e. all result mentioned below are either for transverse voltage in zero magnetic field or for transverse even one.

## 2. Experimental

Two types of samples (A and B) were studied. All the samples were in the form of stripes used for ordinary resistivity measurement with the current leads attached to the ends and point potential contacts along the edges of the sample. Sample B has been prepared by multiple-target sputtering on c-oriented sapphire substrate [3]. The thickness of MgB$_2$ was about 160 nm and it was covered by AlN protection layer to prevent degradation. Voltage and current leads have been attached to the sample by ultrasonic method. The critical temperature T$_c$ of B sample was slightly below 30 K with width of transition region about 0.2 K. Samples of A group were prepared by deposition on silicon (100) single crystal substrate [4]. The T$_c$ was around 27 K and the transition was wider that in B samples (approx. 5 K). The reason for using these samples with broad transition was the possibility to study in detail the transition region. The samples were formed by standard photolithography into shape suitable for six point measurement. Contacts have been made by applying In droplets in the contact area and then soldering leads to them.

Disturbing voltage signals (thermoelectric etc.) have been eliminated by switching off and/or reversing the transport current. The misalignment of transverse contacts was corrected for by measuring the transverse voltage in the regime where no transverse voltage should appear, i.e. well above the transition temperature where the sample is in the normal state.

## 3. Results and discussion

The dependences of the transverse voltage on magnetic field are shown in figures 1 and 2 for the sample A and B, respectively. Increasing magnetic field shifts T$_c$ to slightly lower value, yet the influence on maximum value of transverse voltage is more pronounced. In zero external magnetic field the non-zero transverse voltage is observed. Its value reaches maximum close to T$_c$ and far away from T$_c$, both below and above no such voltage has been detected, i.e. both in superconducting and normal state it is zero. The suppression of the peak with increasing magnetic field is shown for both samples in figures 3 and 4. It seems that the transverse voltage is the continuation of non-zero transverse voltage in zero magnetic field and that this suppression is exponential but more data are needed. It is clear from these figures that the existence of the non-zero transverse voltage is not a result of magnetic field generated by transport current through the sample. Moreover such magnetic field calculated for the current involved is in the range of few μT which is probably too small to influence the observed voltage. Figures 1 and 2 also show the comparison of the temperature dependence of transverse voltage with the derivative of the longitudinal voltage. It is visible that this derivative copies the transverse voltage peak. This is true not only in zero magnetic field but also in non-zero external magnetic field.

The non-zero transverse voltage in zero external magnetic fields has been observed also in other superconducting materials. First observation was made by Francavilla et al. [5] on YBaCuO, for others see for example [2]. Several possible explanations have been suggested. Francavilla et al. used for explanation of their results Glazmann model [6] based on the annihilation of vortices and antivortices created by transport current. Recently [7] it has been shown that this model is not adequate, thanks to the vortex-antivortex symmetry in this system. However a non zero voltage can be observed when induced vortices are moving under influence of any force which violates vortex-antivortex symmetry, for example the guiding force.

But vortices can be generated in the sample in the absence of the external magnetic field also by another way. In superconductors, vortex-antivortex pairs may be excited as thermal fluctuations [8]. Spontaneous creation of the free vortices and antivortices has been considered as a result of thermally activated dissociation of vortex-antivortex pair at Kosterlitz-Thouless temperature T$_{KT}$ [9]. These free vortices and antivortices can move under the influence of an external electric field and cause thus a dissipation which can in principle lead to the appearance of the transverse voltage. Nevertheless there should again exist some force which will drive vortices in the way to contribute to non-zero transverse electric field. One should have in mind, however, that Kosterlitz-Thouless transition occurs in the system of the universality class of the two–dimensional XY model and it is questionable whether MgB$_2$ belongs to such materials.

It has been suggested [10] that in HTS the time reversal symmetry is broken spontaneously in the absence of an applied magnetic field The violation of the time reversal symmetry and of the two-dimensional reflection symmetry (parity) can be a consequence of the fractional statistics of the quasiparticles [11]. In [12[ and [13] the violation is assumed to be a bulk property and it is attributed either to the presence of d$_{x^2+y^2}$+id$_{xy}$ states or to the formation of Π junction. The temperature below which these symmetries are broken is expected to be in the vicinity of the critical temperature. The violation would lead to the appearance of an antisymmetric contribution to the resistivity tensor even in the absence of the applied magnetic field. So the breakdown of time reversal symmetry can lead to the appearance of observed effect. The breakdown should lead also to the violation of the reciprocity theorem. An attempt to prove violation of reciprocity theorem was made on YBaCuO thin film in [14]. However they did not see any such behaviour. But the violation of the reciprocity theorem has been recently confirmed [7]. One of the most interesting consequences of broken T and P symmetry is the prediction of the existence of an intrinsic orbital moment leading to the spontaneous appearance of macroscopic magnetization at the superconducting transition temperature. Such moment was really observed on epitaxial thin films of YBaCuO [12]. Nothing similar was however observed on other superconducting materials where broken reversal symmetry and non-zero transverse voltage has been detected.

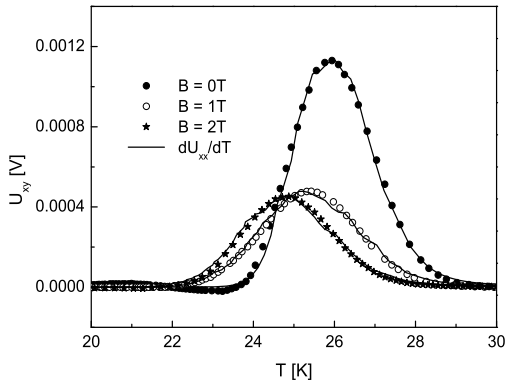

Figure 1: Temperature dependence of $U_{xy}$ as a function of magnetic field compared with normalized derivatives of $U_{xx}$ (sample A)

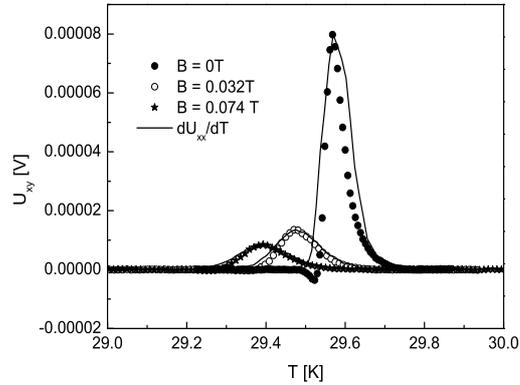

Figure 2: Temperature dependence of $U_{xy}$ as a function of magnetic field compared with normalized derivatives of $U_{xx}$ (sample B)

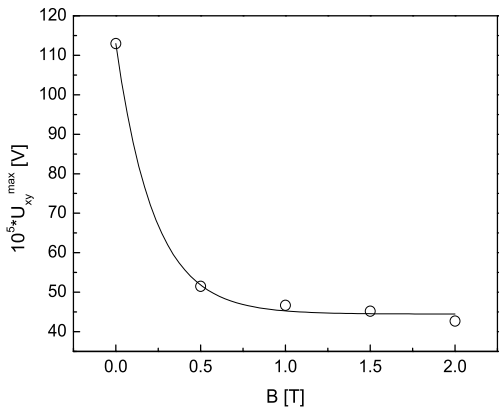

Figure 3: Suppression of the $U_{xy}$ peak by magnetic field (sample A – full line serves as guide for eye only)

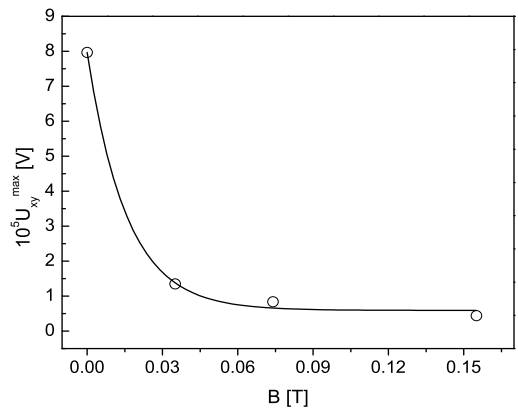

Figure 4: Suppression of the $U_{xy}$ peak by magnetic field (sample B - full line serves as guide for eye only)

As mentioned above, the transverse voltage which is even function of the direction of magnetic field is the continuation of the non-zero voltage observed in zero external field. With increasing magnetic field the even voltage is suppressed. Usually accepted explanation of the even effect is the existence of guided motion of vortices when vortices move in the channels. With increasing magnetic field the Lorentz force can overcome the force which holds vortices in the channel, vortices will move in the direction of driving forces and the even effect disappears.

Figs. 1 and 2 also show that there is a close relation between observed transverse voltage and the temperature derivative of longitudinal one. This can be written in terms of longitudinal and transverse resistivities $\rho_{xx}$ and $\rho_{xy}$, respectively, in the form

$$\rho_{xy} = A \frac{d}{dT} \rho_{xx} \tag{1}$$

where A describes the scaling. The increasing magnetic field does not violate this relation.

If one accepts that all the observed effects are a result of guided motion of vortices a following explanation offers. When the vortex is moving in the channel nonparallel to the driving force, the following equation should be solved [1]

$$\alpha\Phi (\mathbf{v} \times \mathbf{z}) \cdot \mathbf{n} + \eta \mathbf{v} \cdot \mathbf{n} = \Phi(\mathbf{j} \times \mathbf{z}) \cdot \mathbf{n} \tag{2}$$

where $\alpha$ is the Magnus force coefficient, $\Phi$ the magnetic flux quantum, v vortex velocity, z the unit vector in the direction of the magnetic field, $\eta$ viscosity coefficient and n the unit vector in the direction of the channel [1].

The electric field is then given by

$$\mathbf{E} = -\frac{B\Phi}{\eta}[(\mathbf{j} \times \mathbf{z}) \mathbf{n}] \mathbf{n} \times \mathbf{z} \tag{3}$$

It follows from this equation that any component of electric field induced by the motion of vortex in the channel is an even function with respect to the direction of the magnetic field. Taking j = ( j,0,0) and B = (0,0,B) one gets for the electric field component

$$E_{xx} = \frac{jB\Phi}{\eta} \sin^2 \beta \qquad\qquad E_{xy} = -\frac{jB\Phi}{\eta} \sin 2\beta$$

where $\beta$ stands for the angle between direction of the transport current and the direction of the channel. From it follows the relation (1) where $A(T) = \frac{B\Phi}{\eta}\frac{d\beta}{dT}$. The term $\frac{d\beta}{dT}$ describes the thermal dependence of the guide angle. With increasing temperature the vortex can overcome the potential barriers of the channel due to the thermal excitation and the decrease of $\beta$ can be observed. Although the suggested explanation can in principle explain the scaling (1) one should have in mind that mentioned description is only very rough based on the simple model and more theoretical work is needed for clarification of this point.

**4. Conclusions**

The existence of non-zero transverse voltage has been verified for $MgB_2$ around temperature of superconducting transition. Measurement in magnetic field shows close connection between transverse voltage in zero magnetic field and transverse voltage which is an even function of direction of magnetic field. The observed results are discussed, suggesting several mechanism (violation of time reversal symmetry, guided motion etc). The most promising explanation seems the guided motion of vortices created in zero magnetic field by thermally activated dissociation of vortex-antivortex pairs. Acceptance of this model can in principle explain the observed relation between transverse and longitudinal resistivities, which is also valid in magnetic fields where the even transverse voltage can be observed.

*Acknowledgements*   This work was partially supported by GACR projects 203/05/0144 and 202/05/0173.